\documentclass[aps,prl,10pt]{revtex4-1}
\usepackage{graphicx}
\usepackage{color}
\begin{document}
\title{Self-consistent plasma wake field interaction in quantum beam transport}
\author{F. Tanjia}
\affiliation{{\small Dipartimento di Scienze Fisiche, Universi$\grave{a}$ di Napoli ``Federico II" and INFN, Napoli, Italy}}

\begin{abstract}
A review of a recent theoretical investigation of the quantum transverse beam motion that has been developed in terms of a coupled system of nonlinear spinorial equations is carried out. This is done assuming that a relativistic electron/positron beam is traveling through a plasma along an external magnetic field in overdense condition. Three different nonliner regimes of this analysis are considered: the strictly local regime, where the beam spot size is much greater than the plasma wavelength (then, the spinorial system is reduced to a 2D cubic nonlinear Schr\"{o}dinger equation in the form of Gross-Pitaevskii equation); the moderately nonlocal regime where the beam spot size is comparable to the plasma wavelength; and, finally, the strongly nonlocal regime where the beam spot size is much smaller than the plasma wavelength. In all the cases the existence of quantum beam vortices and ring solitons is demonstrated both analytically and numerically.
\end{abstract}

\maketitle

\section{Introduction}\label{Introduction}
Plasma is an attractive medium for particle acceleration \cite{Fedele1990} because it can sustain very high electric and magnetic fields \cite{Dawson1959a, Dawson1959b}. The physical processes involved in the generation of large amplitude fields are relevant for many scientific and technological applications. In particular, the studies of plasma-based charged particle acceleration constitute one of the most important areas of research in laboratory, space and astrophysical plasmas. A plasma has no electrical breakdown limit like conventional accelerating structures, which are limited to a maximum field strength of less than 1 MV/cm. A plasma supports longitudinal plasma waves \cite{Dawson1959a, Dawson1959b}, in which the plasma electrons oscillate back and forth at the plasma frequency $\omega_p$ irrespective of the wavelength. Therefore, these waves can have arbitrary phase velocity $v_p$. In the case of relativistic plasma waves, $v_p$ is almost of the order of the light speed $c$. Particle acceleration by relativistic plasma waves has gained a lot of interest lately due to both the rapid advances in laser technology and the development of compact terawatt and petawatt laser systems with ultra-high intensities ($\geq 10^{20}$ W cm$^{-2}$), modest energies (a few or tens joules) and ultra-short pulse durations (a few femto-second).
Alternatively, instead of using lasers, short relativistic charged-
particle beams can also excite relativistic large amplitude plasma waves.
The study of relativistic charged particle beam dynamics in plasmas has increased gradually in connection with the richness of nonlinear and collective effects induced by the propagation of very intense and very short charged particle bunches. Intense charged particle bunches propagating in a plasma excite ultra-intense plasma fields. These fields provide, in turn, intense acceleration of particles. The typical charged particle beam-driven plasma wave excitation is the well known Plasma Wake Field (PWF) excitation \cite{Chen1985, Rosenzweig1988, Rosenzweig1991}. In the PWF excitation, a charged particle beam or bunch (driver) travels in a neutral plasma with ions as background.
If the driver is a relativistic electron beam, then the Coulomb force of the beam's space charge expels plasma electrons, which rush back in after the beam and produce a large amplitude plasma wave behind the driver. This plasma wave oscillates at the electron plasma frequency and follows the driver much the same way water wakes follow a fast boat (plasma wake) \cite{Chen1985}. Its phase velocity is therefore equal to the driver velocity, and is almost independent of the plasma density. The electromagnetic (e.m.) field associated to the wake (wake field) has transverse as well as longitudinal  components. Thus, a test particle experiences the effects of both the transverse (focusing/defocusing) and the longitudinal (acceleration/deceleration) components of the wake field. Depending on the regimes, the test particle can be the one of a secondary beam externally injected in phase locking with the wake (driven beam) or belonging to the driver. In the last circumstance, the driver experiences the effects of the wake field that itself produced. Taking into account all together the effects on each particle of the driver we can describe the collective self-interaction of the driver with the plasma.
In the case of positrons, the electrons of the plasma background are pulled in by the driver which overshoot and set up the plasma oscillation. Then, it is easily seen that we can provide PWF interaction for a relativistic positron beam in a way fully similar to the one described above for an electron beam.

The aim of this paper is to present a review of the results of a recently developed theory on the transverse effects due to the interaction of relativistic charged particle beams with a magnetized plasma via PWF excitation in the quantum regime \cite{Fedele2012a, Fedele2012b}. In section \ref{PWF}, the theory of PWF excitation is briefly presented by assuming a fluid model of magnetized plasma (Lorentz-Maxwell system of equations) and taking into account the \textit{long beam limit}. In section \ref{Paraxial}, the governing coupled system of equations are presented in cylindrical symmetry by taking into account the individual quantum nature of the particles but disregarding the collective one (overlapping of the wave function) according to Hartree's mean field procedure. Then, in sections \ref{Local nonlinear}, \ref{Moderate nonlocal}, and \ref{Strictly nonlocal}, the analytical and numerical results of the system for the nonlinear case in strictly local, moderately nonlocal, and strictly nonlocal regimes, respectively, are presented briefly. Finally, the conclusions and remarks are presented in section \ref{Conclusion}.

\section{Self-consistent PWF interaction in a magnetized plasma}\label{PWF}
The starting point for most analysis of nonlinear wave propagation phenomena in plasma is the fluid model. It consists of Lorentz fluid equations of motion together with Maxwell's equations - the so called \textit{Lorentz-Maxwell system}. We assume that the plasma, whose unperturbed density is $n_0$, is cold, collisionless and magnetized due to the presence of an external uniform magnetic field $\mathbf{B}_0$, oriented along the $z-$direction ($\mathbf{B}_0=\hat{z}B_0$). In addition, we assume that a relativistic electron/positron beam, whose unperturbed density is $n_b$, is traveling along the magnetic field direction, in the overdense condition ($n_{0}\gg n_{b}$) and with the unperturbed velocity $\beta c \hat{z}$ ($\beta\simeq 1$). Finally, the ions are assumed to be immobile and forming a uniform background of density $n_0$.

Hereafter, ignoring the longitudinal beam dynamics, we introduce small perturbations of all the physical quantities, such as plasma electron density, plasma electron velocity, the transverse perturbation components of the beam velocity, electric and magnetic field. We express the fields $\mathbf{E}$ and $\mathbf{B}$ in terms of the four-potential ($\mathbf{A}, \phi$) and apply the Lorentz gauge, viz., $\nabla\cdot\mathbf{A}+\left(1/c\right)\partial\phi/\partial t=0$. Next, we transform all the fluid equations to the beam co-moving frame $\xi=z-\beta ct\simeq z-ct$ and split all the quantities as well as gradient operator $\nabla$ into longitudinal and transverse components. In the \textit{long beam limit}, i. e., if the beam length is much greater than the plasma wavelength, finally, we solve the system of equations, to get (for details, see refs. \cite{Fedele2012a, Fedele2012b})
\begin{equation}
\left(\nabla^2_\perp-\frac{k^4_{pe}}{k_{uh}^2}\right)U_w=\frac{k^4_{pe}}{k_{uh}^2}\frac{\rho_b}{n_0\gamma_0}\,,\label{a7-2}\
\end{equation}
where $U_w(\textbf{r}_\perp,\xi)=-q (A_{1z}-\phi_1 )/m_0\gamma_0c^2$ is the dimensionless wake potential, while $A_{1z}(\textbf{r}_\perp,\xi)$ and $\phi_1(\textbf{r}_\perp,\xi$) are the longitudinal component of the vector potential and scalar potential perturbations, respectively, $\rho_b$ is the beam number density, $\gamma_0$ is the relativistic factor, $e$ is the magnitude of the electron charge, and $m_0$ is the rest mass of the electron or positron. Here, $k_{uh}=\omega_{uh}/c$, $k_{pe}=\omega_{pe}/c$, $k_{ce}=\omega_{ce}/c$, $\omega_{uh}=(\omega^2_{pe}+\omega^2_{ce})^{1/2}$ is the upper hybrid frequency, $\omega_{pe}=(4\pi n_0e^2/m_0)^{1/2}$ is the {electron plasma} frequency and $\omega_{ce}=-eB_0/m_0c$ is the electron cyclotron frequency.

\section{The spinorial equations in quantum paraxial diffraction}\label{Paraxial}
\subsection{Paraxial beam}
Paraxial approximation is a frequently used approximation, essentially assuming small angular deviations of the particle trajectories with respect to the beam propagation direction. This concept is fully similar to the one encountered in e.m. optics when we assume that in an e.m. beam (such as a laser) the light rays slightly deviate from the beam axis while propagating. Given the well known analogy between the geometrical e.m. optics and classical electron optics \cite{Sturrock1995}, a common language is now adopted for both disciplines and similar terminologies are used to describe analogous situations. The general statement from which we recover such a generalized description of geometrical optics is the analogy between Fermat principle (e.m. geometrical optics) and Hamilton principle (classical mechanics). Later on, as it is well known, the analogy has been successfully extended to the wave context between e.m. wave optics and wave mechanics from which the modern formulation of the quantum mechanics has been originated.

A `quantum' description to the charged particle beam transport \cite{Fedele2012a,Fedele2012b} requires a quantization procedure in which the physical quantities associated with the single-particle are interpreted as operators according to the Bohr quantization procedure. In the next sections, we discuss this procedure and describe the paraxial approximation of the charged particle beam transport. We introduce the concept of \textit{quantum paraxial diffraction} for a charged particle beam taking into account the individual quantum nature of the particles (single-particle uncertainty principle and spin). To present the concept in a simple way, we disregard the overlapping of the particle wave functions.

\subsection{Relativistic single particle Hamiltonian}
Let us consider the relativistic classical Hamiltonian of a single particle of the charged beam in the presence of the four-potential ($A,\phi$)\begin{equation}
H=c\left[(\mathbf{p}-\frac{q}{c}\mathbf{A})^2+m^2_0c^2\right]^
{1/2}+q\phi, \label{h1}\
\end{equation}
where $\mathbf{p}$ is the canonical momentum of a single particle of the beam. We split $\mathbf{p}$ and $\mathbf{A}$ into the longitudinal and transverse components, viz., $\mathbf{p} = \hat{z}p_z + \mathbf{p}_\perp$, $\mathbf{A}= \hat{z}A_z + \mathbf{A}_\perp$. We are interested in the transverse dynamics of the beam, so we set $p_z = m_0\gamma_0\beta c\equiv p_0$ for the entire duration of the beam motion. In order to be consistent with the paraxial approximation, the beam particles may have a relativistic motion along $z$ (longitudinal motion) but the transverse component of the single particle motion must be non relativistic. This implies that the non relativistic expansion of the Hamiltonian in transverse components is possible, i.e. $H$ can be expanded in $\mathbf{P}_\perp-\mathbf{A}_\perp$, up to the second power. Then, similarly to the previous section, we introduce small perturbations for the quantities $\mathbf{A}_z$, $\mathbf{A}_\perp$, and $\phi$, viz., $\mathbf{A}_z =\mathbf{A}_{1z}$, $\mathbf{A}_\perp =\mathbf{A}_0 +\mathbf{A}_{1\perp}$, and $\phi =\phi_1$. Then, the combined action of the non relativistic expansion of the Hamiltonian (\ref{h1}) and its linearization lead to an effective dimensionless classical Hamiltonian \cite{Fedele2012a,Fedele2012b}.

\subsection{Quantization procedure}
Now, we take into account the quantum nature of the beam particles. First of all, let us consider the quantum nature of each individual particle. By using the language of the electron optics, at each $z$, the uncertainty relation between the position and the momentum, which holds for a single electron/positron, is translated as the uncertainty relation between the position and the slope of the individual electron ray in the transverse plane. This way, the standard paraxial ray picture of a charged particle beam is implemented by the quantum uncertainty. It may be expected that, due to the \textit{quantum paraxial} behavior, the beam cannot have an arbitrarily small spot size. In fact, it can be reduced to a certain minimum value, according to the uncertainty principle.

Taking into account the spin and the uncertainty relation of each particles, we can now describe the transverse beam dynamics by correcting the paraxial geometric electron optics with the paraxial wave electron optics, given in terms of a single-particle wave function \cite{Fedele2012a,Fedele2012b}. In principle, the effective quantum description of the transverse motion of the beam, containing $N$ charged particles, can be provided by the system of $N$ two-dimensional Schr\"{o}dinger equations whose Hamilton operator corresponds to the classical Hamiltonian $H$ defined in Eq. (\ref{h1}). According to the prescriptions of quantum mechanics, to write the appropriate Schr\"{o}dinger equation for a single particle of the beam, which includes also the spin, we introduce the Bohr correspondence rules into the classical Hamiltonian given in it's dimensionless form,  where $\hbar$ is replaced by $\epsilon\equiv\hbar/m_0\gamma_0c=\lambda_c/\gamma_0$, where $\lambda_c$ is the Compton wavelength. Therefore, the quantity $\epsilon$ plays the role of the relativistic Compton wavelength. Note that, here the Hamiltonian operator $\hat{\mathcal{H}}$ includes also the potential energy operator $\hat{\mathcal{U}}_M=-\hat{\mu}_s\cdot \mathbf{B}_0 /m_0\gamma_0c^2$ associated with the interaction between the intrinsic magnetic moment of spin and the magnetic field. Here $\hat{\mu}_s=q\epsilon\hat{s}$, where $\hat{s}$ is the electron/positron spin operator. Therefore, $\hat{\mathcal{U}}_M = \hat{\mathcal{U}}_M(\hat{s}_z)$, where $\hat{s}_z$ is the {projection} of $\hat{s}$ along $z$. Since the spin of an elementary particle is a purely quantum property, there is no classical term in the classical Hamiltonian (\ref{h1}) that corresponds to $\hat{\mathcal{U}}_M$. Thus, the quantized dimensionless Hamiltonian has the form (for details see \cite{Fedele2012a,Fedele2012b})
\begin{equation}
\hat{\mathcal{H}}=\frac{1}{2}\hat{\mathcal{P}}^2_{\perp}+\frac{1}{2}k_c\hat{z}\cdot(\mathbf{r}_\perp\times\hat{\mathbf{\mathcal{P}}}_{\perp})
+U_w(\textbf{r}_\perp,\xi)+\frac{1}{2}Kr^2_\perp +\hat{\mathcal{U}}_M(\hat{s}_z)\,, \label{h3}\
\end{equation}
where $\mathcal{H}=\Delta H/H_0=(H-H_0)/H_0$ is the effective dimensionless Hamiltonian, $H_0=c(p^2_0+m^2_0c^2) ^{1/2}=m_0\gamma_0c^2$ is the unperturbed
energy of the single particle, $\mathbf{\mathcal{P}}_{\perp}=\mathbf{p}_{\perp}/m_0\gamma_0c$ is the dimensionless perpendicular momentum, $k_c=-qB_0/m_0
\gamma_0c^2$, and $K\equiv\left(k_c/2\right)^2=\omega_{ce}^2/4\gamma_0^2c^2$.  Note that, the quantity  $\hat{z}\cdot(\mathbf{r}_\perp\times\hat{\mathbf{\mathcal{P}}}_{\perp})$ is the longitudinal component of the orbital angular momentum due to the presence of external magnetic field, and $U_w(\textbf{r}_\perp,\xi)+Kr_\perp^2/2$ plays the role
of a multiplication operator (the second term of the sum being the focusing potential well provided by the dipole magnetic field $\mathbf{B}_0$).

\subsection{The governing coupled system of equations}
On the basis of these considerations, it is easy to see that the application of the Hartree’s mean field procedure allows us to reduce our description to finding a \textit{single-particle spinor} $\overrightarrow{\Psi}=(\Psi_s)$, whose spatio-temporal evolution is governed by two-dimensional Schr\"{o}dinger-like equations, where $s$ is ranging in the set $\{-1/2,1/2\}$ of eigenvalues of $\hat{s}_z$ corresponding to the ``spin down" and ``spin up" states, respectively. On the other hand, since the beam is composed of only two kinds of spin states of electrons/positrons, according to the Hartree’s approach, the beam density is given by $\rho_b=\left(N/2\sigma_z\right)|\overrightarrow{\Psi}|^2=\left(N/2\sigma_z\right)|\Psi_{-1/2}|^2+
|\Psi_{1/2}|^2$, where we have assumed that statistically the beam population is equipartite into “down” and “up” spin states and $\sigma_z$ is the beam length. Here, we also assumed that each $\Psi_s$ is normalized. Thus, we can conclude that the self-consistent beam-plasma interaction that accounts for the PWF excitations and the quantum paraxial beam diffraction is governed by the following coupled system of equations (for details see \cite{Fedele2012a,Fedele2012b})
\begin{eqnarray}
&&i\epsilon\frac{\partial\overrightarrow{\Psi}}{\partial \xi}=-\frac{\epsilon^2}{2}\nabla_\perp^2\overrightarrow{\Psi}-\frac{i\epsilon k_c}{2}\hat{z}\cdot(\mathbf{r}_\perp\times\nabla_\perp)\overrightarrow{\Psi}
+\left(U_w+\frac{1}{2}Kr^2_\perp\right)\overrightarrow{\Psi}+\epsilon k_c\hat{s}_z\cdot\overrightarrow{\Psi}, \label{h4}\\
&&\left(\nabla^2_\perp-\frac{k^4_{pe}}{k_{uh}^2}\right)U_w=\frac{k^4_{pe}}{k^2_{uh}}\frac{N}{2n_0\gamma_0\sigma_z}|\overrightarrow{\Psi}|^2. \label{h4-b}
\end{eqnarray}
In cylindrical coordinates, for the case of the solutions of Eq. (\ref{h4}) with the form $\Psi_s(r_\perp,\xi,\varphi)=\psi_m(r_\perp,\xi)\exp\{i[m\varphi-(k_c/2) (m+2s )\xi ] \}$, where $m$ is integer, and $r_\perp$ and $\varphi$ are the polar coordinates in the transverse plane, $U_w[|\psi_m|^2]$ becomes cylindrically symmetric and the system can be cast in the following spin-independent form
\begin{eqnarray}
&&i\epsilon\frac{\partial \psi_m}{\partial\xi}=-\frac{\epsilon^2}{2}\frac{1}{r_\perp}\frac{\partial}{\partial r_\perp}\left(r_\perp\frac{\partial \psi_m}{\partial r_\perp}\right)+U_w\psi_m+\left(\frac{1}{2}Kr^2_\perp+\frac{m^2\epsilon^2}{2r^2_\perp}\right)\psi_m, \label{c3}\\
&&\frac{1}{r_\perp}\frac{\partial}{\partial r_\perp}\left(r_\perp\frac{\partial U_w}{\partial r_\perp}\right)-\frac{k^4_{pe}}{k^2_{uh}}\,U_w=\frac{k^4_{pe}}{k^2_{uh}}\frac{N}{n_0\gamma_0\sigma_z}\,|\psi_m|^2. \label{c4}\
\end{eqnarray}
This is the pair of governing equations that have been solved and studied, both in linear and nonlinear regime \cite{Fedele2012a,Fedele2012b,Jovanovic2012b}, in order to show the
existence of nonlocal coherent structures associated with the vortex states of the beam. In the next sections, we briefly describe the results of the nonlinear regime for different nonlocality.
\begin{figure}
\includegraphics[width=13cm]{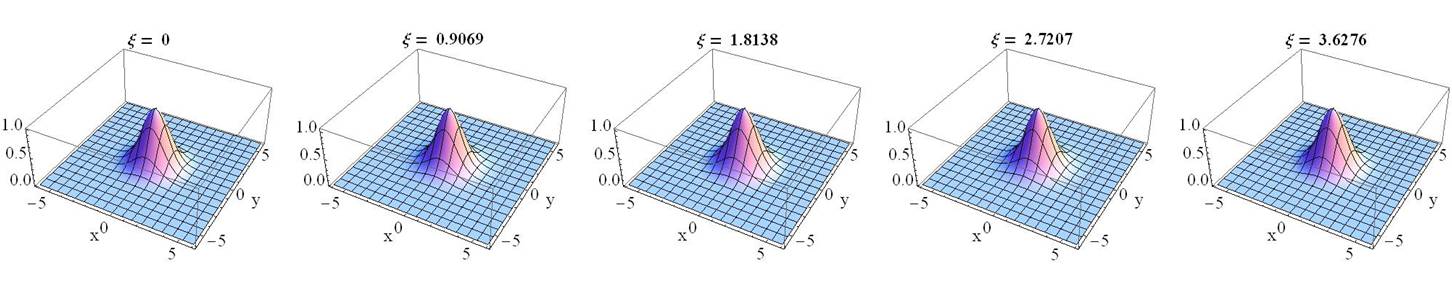}
\caption{3D plots of $|\psi_m|^2$ as a function of $x$ and $y$ for different values of dimensionless time $\xi$ for $m = 0$, $K_b = 0.75$, $\delta = 0.5$, $\tilde{A}_m =0.75$(self equilibrium), as reported in \cite{Fedele2012b}.}
\end{figure}
\begin{figure}
\includegraphics[width=13cm]{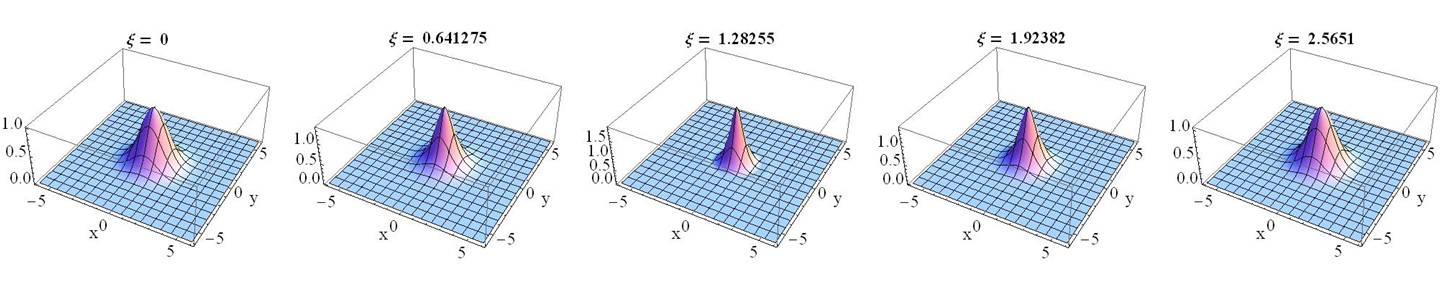}
\caption{3D plots of $|\psi_m|^2$ as a function of $x$ and $y$ for different values of dimensionless time $\xi$ for $m = 0$, $K_b = 1.5$, $\delta = 1.0$, $\tilde{A}_m =1.0$ (focusing - defocusing), as reported in \cite{Fedele2012b}.}
\end{figure}
\begin{figure}
\includegraphics[width=13cm]{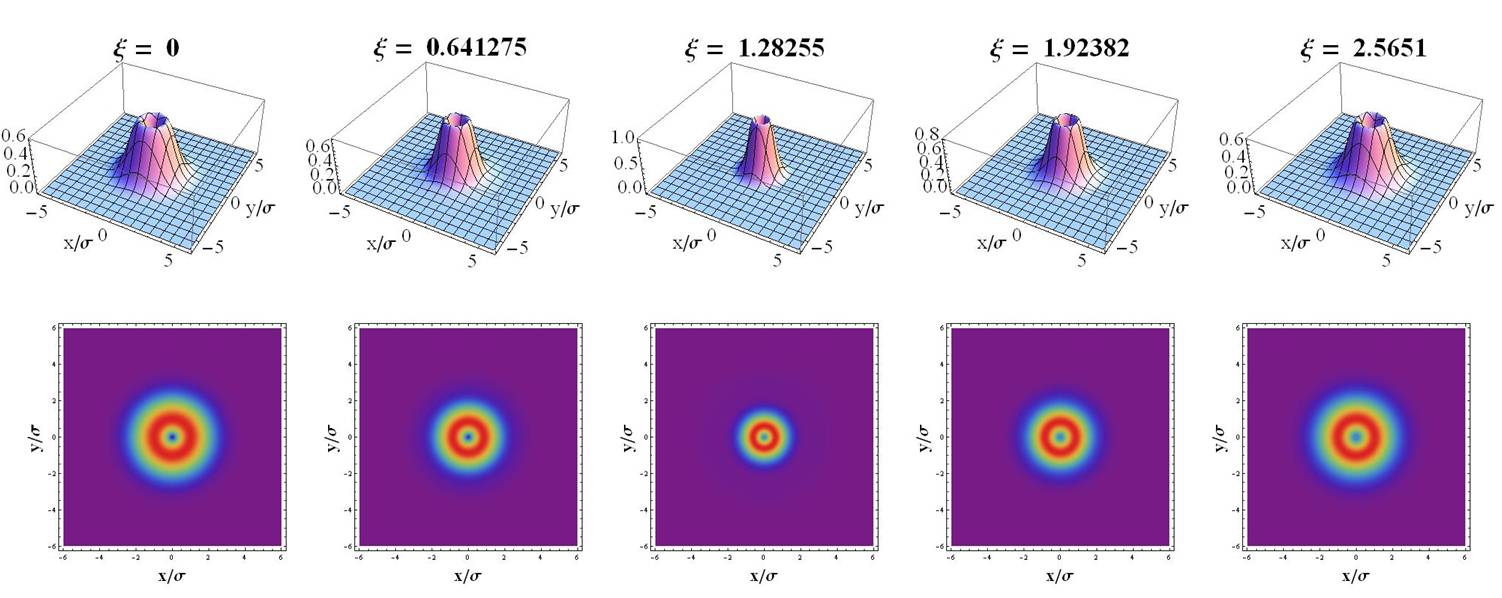}
\caption{3D plots of $|\psi_m|^2$ as a function of $x$ and $y$ for different values of dimensionless time $\xi$ for $m = 1$, $K_b = 1.5$, $\delta = 3.5$, $\tilde{A}_m =2.0625$ (focusing - defocusing), as reported in \cite{Fedele2012b}.}
\end{figure}
\begin{figure}
\includegraphics[width=13cm]{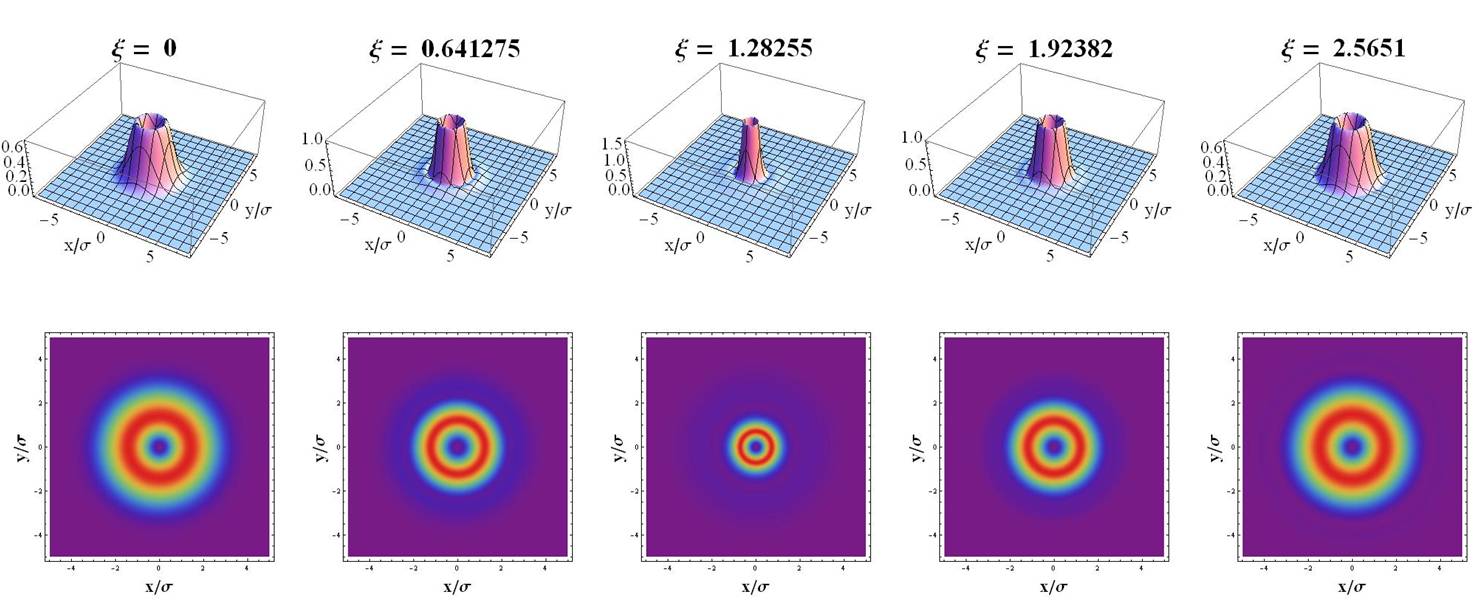}
\caption{3D plots of $|\psi_m|^2$ as a function of $x$ and $y$ for different values of dimensionless time $\xi$ for $m = 2$, $K_b = 1.5$, $\delta = 5.0$, $\tilde{A}_m =5.625$ (focusing - defocusing), as reported in \cite{Fedele2012b}.}
\end{figure}
\begin{figure}
\begin{center}
\includegraphics[width=10cm]{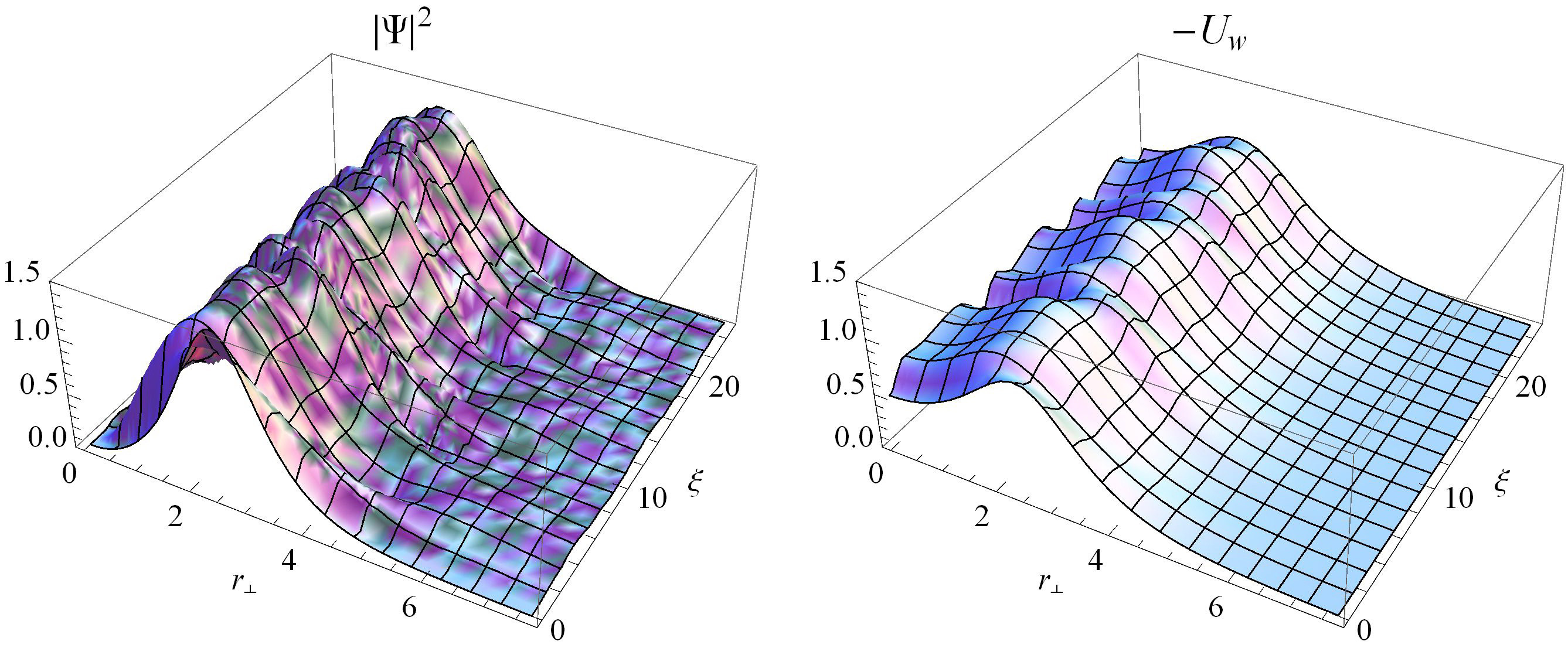}
\caption{Ring soliton of a moderate  beam. The evolution of the wave function (left) and of the wake potential (right), as reported in \cite{Fedele2012b}. The initial wave function was adopted as a slab soliton, with the amplitude $a = 1.5$ and with the velocity equal to zero, $b=0$. The azimuthal wavenumber is adopted as $m=2$ and the coefficient of the restoring force is $K=0.1$.} \label{fig:1-moderately_nonlocal_stationary}
\end{center}
\end{figure}
\begin{figure}
\begin{center}
\includegraphics[width=10cm]{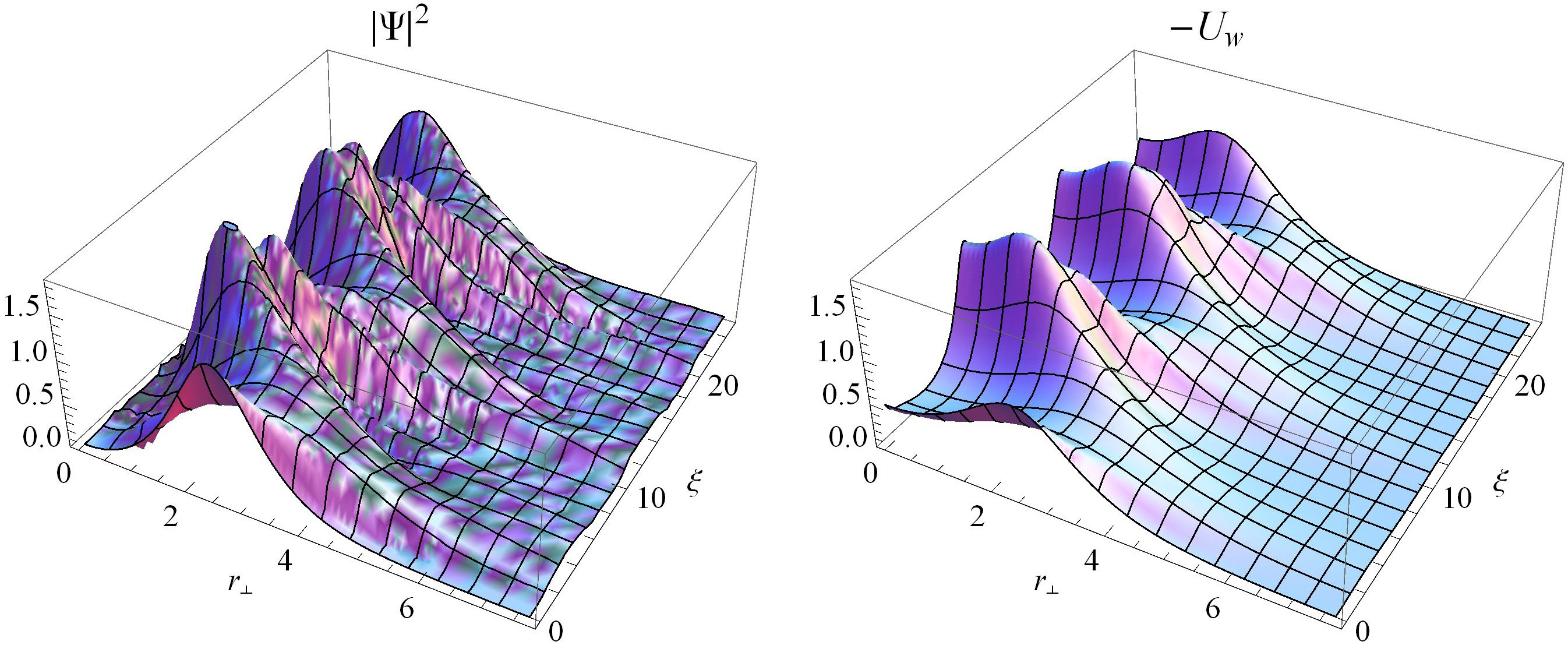}
\caption{The evolution of the wave function (left) and of the wake potential (right), as reported in \cite{Fedele2012b}. All parameters are the same as in Fig. 7, except a finite initial velocity of the structure, $b=0.7$.} \label{fig:2-moderately_nonlocal_bouncing}
\end{center}
\end{figure}
\begin{figure}
\begin{center}
\includegraphics[width=5cm]{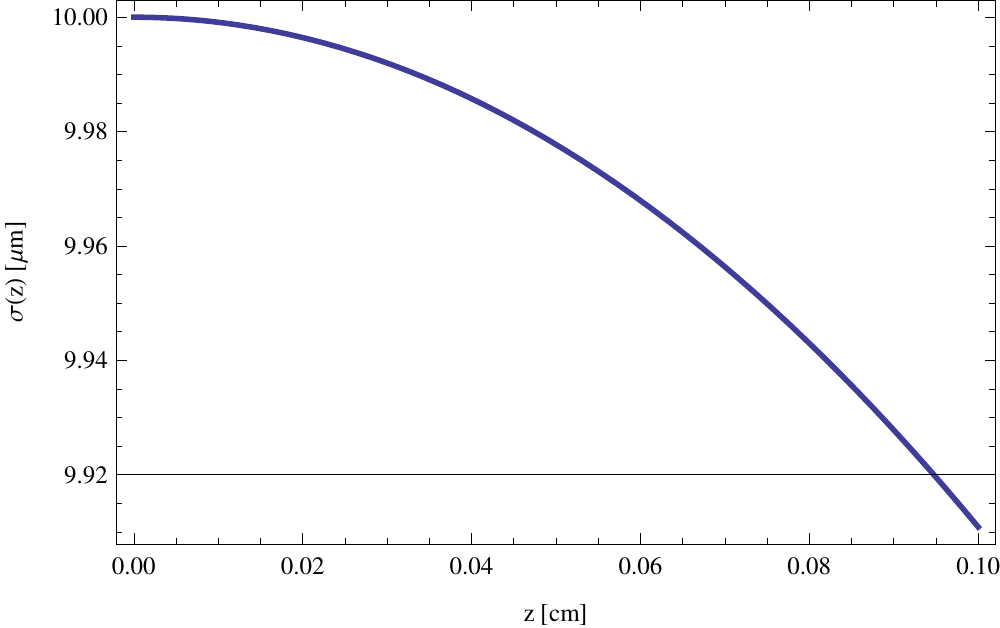}
\includegraphics[width=5cm]{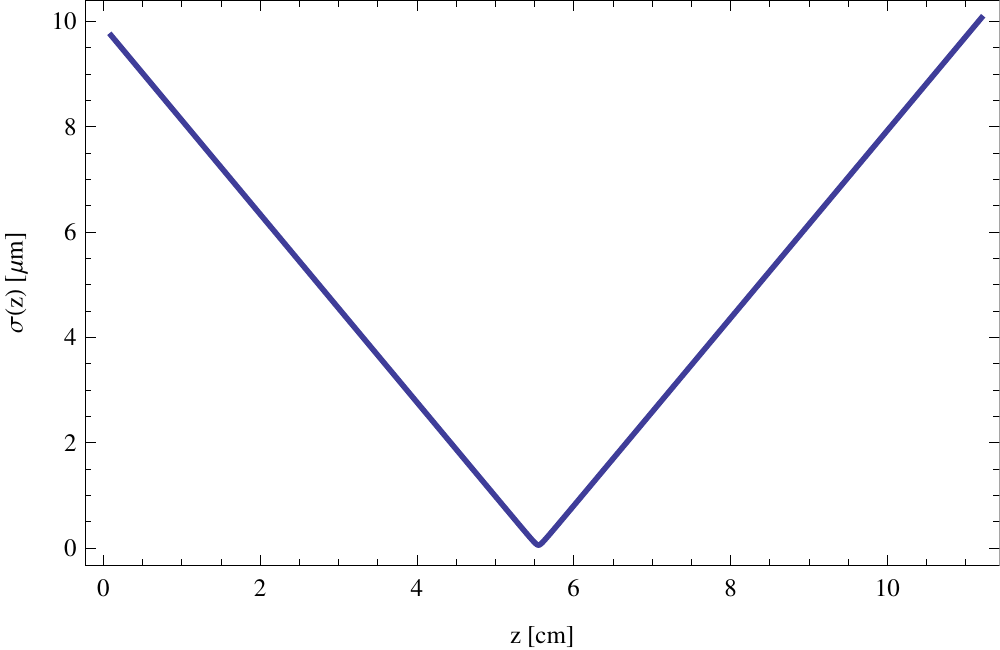}
\caption{Left: Variation of $\sigma$ as a function of $z$ [in cm] inside the lens with the initial condition of $\sigma(0)=\sigma_0$ and $\sigma'(0)=0$, as reported in \cite{Tanjia2013b}. For a lens of thickness $l=1$ mm, it has been found that $\sigma(l)=9.91\,\mu$m and $f=5.55$ cm. Right: Variation of $\sigma$ as a function of $z$ in the vacuum with initial condition $\sigma(0) = \sigma(l)$ and $\sigma'(0) = \sigma(l)/ \rho(l)$. $\sigma$ experiences strong focusing due to the dominance of the magnetic field term $K$. When $\sigma$ becomes very small near the final focus, the space charge effects becomes dominant compared magnetic field and it blows up. The beam spot size in the interaction point is $\sigma* = 56$ nm.}\label{f7-8}
\end{center}
\end{figure}
\section{Strictly local nonlinear regime}\label{Local nonlinear}
In this section, we assume that the relation between the wake potential and the beam density is strictly local. This is obtained by assuming in Eq. (\ref{c3}) that, $\nabla^2_\perp U_w\ll k^4_{pe}/k_{uh}^2U_w$ (strictly local regime) \cite{Fedele2012a,Fedele2012b}. Then we get a 2D cubic nonlinear Schr\"{o}dinger (NLS) equation in the form of the 2D Gross-Pitaevskii equation (GPE) for a Bose Einstein condensate \cite{Gross1961, Pitaevskii1961} in the presence of the cylindrically symmetric external potential well $\left(Kr_\perp^2+m^2\epsilon^2/r_\perp^2\right)/2$, viz.,
{\small\begin{equation}
i\epsilon\frac{\partial \psi_m}{\partial\xi}=-\frac{\epsilon^2}{2}\frac{1}{r_\perp}\frac{\partial}{\partial r_\perp}\left(r_\perp\frac{\partial \psi_m}{\partial r_\perp}\right)-\frac{N}{\sigma_zn_0\gamma_0}\mid\psi_m\mid^2\psi_m+\left(\frac{1}{2}Kr^2_\perp+\frac{m^2\epsilon^2}{2r^2_\perp}\right)\psi_m, \label{s1}\
\end{equation}}
We get the envelope description of the beam by using virial equation as (for details see ref. \cite{Fedele2012b}).
\begin{eqnarray}
&&\frac{d^2\sigma^{2}_{m}}{d\xi^2}+4K\sigma^{2}_{m}
=4\mathcal{A}_m, \label{e4}\\
&&\frac{d\mathcal{A}_m}{d\xi} =0\,,\label{c4-1}
\end{eqnarray}
where
$$\mathcal{A}_m=2\pi\left[\int\frac{\epsilon^2}{2}\left(\left|\frac{\partial\psi_m}{\partial r_\perp}\right|^2+\frac{m^2}{r_\perp^2}\left|\psi_m\right|^2\right)r_\perp dr_\perp - \frac{N}{2\sigma_zn_0\gamma_0}\int\mid\psi_m\mid^4r_\perp dr_\perp\right]+\frac{1}{2}K\sigma^2_m\,,$$
is the constant of motion for the harmonic oscillator equation (\ref{e4}). It can be easily observed from eq. (\ref{e4}) that the matching condition corresponding to the self-equilibrium state of the beam is $\sigma_{m}=\sigma_{m}^{eq}$ is: $K\left(\sigma_{m}^{eq}\right)^2 =\mathcal{A}_m$. Assuming a beam with the initial wave function of Laguerre-gauss form, Eq. (\ref{s1}) has been solved numerically by introducing dimensionless  form (for details see \cite{Fedele2012b}). It is easy to see that such initial condition fixes automatically $\langle r_\perp^2\rangle_{m}(\xi=0)\equiv\sigma_{0m}=\left(m+1\right)\sigma_0$ and $\langle r_\perp^2\rangle'_{m}(\xi=0)\equiv\sigma'_{0m} = \left(m+1\right)\sigma'_0 = 0$, where $\sigma_0$ is the initial beam spot size. The spatio-temporal evolution of $\left|\psi_m\right|^2$ has been investigated for different values of $m$, $K_b$, and $\delta$, at $\xi= 0, 0.25 T, 0.5T, 0.75T, T$, where $K_b=K\sigma_0^4/\epsilon^2$, $\delta=n_b\sigma_0^2/n_0\gamma_0\epsilon^2$, and $T=\pi/\sqrt{K_b}$ is the time period. For $m=0$, when the matching condition of the envelope equation is satisfied, the profile is practically unchanged (see Figure 1). This predicts the existence
of nonlinear coherent states (sometimes called 2D solitons). For $m=1$ and $m=2$, the same self-equilibrium state would possible to observe \cite{Fedele2012b}. Due to the interplay between the strong transverse effects of the plasma wake field (collective and nonlinear effects) and the external magnetic field, envelope oscillations with weak and relatively strong focusing and defocusing have been observed for $m=0$, $m=1$, and $m=2$ as shown in the 3D as well as density plots of Figures 2, 3, and 4, respectively.

\section{Moderately nonlocal nonlinear regime}\label{Moderate nonlocal}
In moderately nonlocal regime, which, in Eq. (\ref{c3}), corresponds to assuming that the beam spot size is comparable to the plasma wavelength, viz., $\nabla^2_\perp U_w\sim k^4_{pe}/k_{uh}^2U_w$. We solved numerically the reduced system of equations by introducing convenient dimensionless form (for details see \cite{Fedele2012b}). First, we looked for a solution whose initial velocity was equal to zero. We adopted $K = 0.1$, $m = 2$, and the following parameters of the initial condition $a = 1.5$, $b = 0$. The numerical solution is displayed in Fig. 5. The solution remained remarkably stable for a long time, $\xi_{max} \geq 25$. During this period, periodic ``breathing" was observed, with a relatively small amplitude, similar to that found in the preceding section, for the local regime.

Next, we looked for a solution with a finite initial velocity and we adopted the same parameters as in Fig. 5, except $b = 0.7$. The solution is displayed in Fig. 6. It also remained stable for a long time, $\xi_{max} \geq 25$, but it exhibited violent oscillations in the radial direction, as well as the `breathing' and the periodic variation of the amplitude.

Using this procedure, we were able to obtain stable solutions only in the weak and moderately nonlocal regimes. In the cases of very narrow initial wave function, the numerical solutions were unstable.

\section{Strongly nonlocal nonlinear regime}\label{Strictly nonlocal}
The theory has been further extended in the strongly nonlocal regime when we can take the approximation $\nabla_\perp^2U_w\gg \left(k_{pe}^4/k_{uh}^2\right)U_w$ \cite{Jovanovic2012b}. This means that the attraction among the beam particles takes place in such a way that the beam may exhibit a strong self-focusing. Following a procedure similar to one developed earlier in nonlinear optics and improved by an appropriate model for the effects of the finite width of the response function, we have found the coherent nonlinear solutions similar to the accessible
solitons in nonlinear optics. However, in contrast to the nonlocal response functions pertinent to the nonlinear optics that are regular, in the case of a relativistic particle beam, the response function possesses a singularity at the beam axis and the analogue of the fundamental two-dimensional optical soliton could not be found in this regime. It has been shown that, besides the stable stationary states, the Hermite-Gauss thin ring solitions in particle beams may feature also the ``breathing” (the oscillations of their width) and “wiggling” (oscillations around the equilibrium position) coherent states that can be identified as the generalized Glauber modes. Such time dependent states may resonantly self-couple with their natural frequency, giving rise to a parametric instability and, eventually, to the destruction of the coherence of the particle beam. The conditions for such resonant coupling were determined and the corresponding growth rates were estimated analytically.
\subsection{Quantum plasma lens}
In the strictly nonlocal regime, very recent work \cite{Tanjia2013b} shows a preliminary investigation to conceive a plasma lens \cite{Chen1987a, Fedele1990,Su1990} in quantum regime as a possible application towards focusing of the beam. The analysis has been carried out in the aberration-less approximation. When the thick plasma medium is replaced by a slab of thickness $l=1$mm, the periodic breathing of the transverse spot size $\sigma$ is no longer existing, rather it shows a slight bending of few fractions of $\mu$m keeping the value of $\sigma$ almost unchanged inside the slab, as shown in Figure 7. When the beam leaves the lens and goes into the vacuum, a strong focusing of $\sigma$ until the focal length $f$ is observed due to the dominance of the magnetic field term $K\sigma$ over the space charge term $\alpha/\sigma$, as shown in Figure 7. When $\sigma$ becomes much smaller reaching $f$, the term $\alpha/\sigma$ becomes dominant over $K\sigma$, and it blows up afterwards. We found the beam spot size at the final focus is $\sigma*\simeq 56$ nm, which is three order smaller than the initial beam spot size. For details see ref. \cite{Tanjia2013b}.

\section{Conclusions}\label{Conclusion}
The recent theoretical investigations of the self-consistent PWF interaction of a charged particle beam with a magnetized plasma has been reviewed. This has been done by taking into account the individual quantum nature of the particle (single particle uncertainty relation and spin) in paraxial approximation. It has been shown that this approach, for all the regimes (i.e., strictly local, moderately nonlocal, and strongly nonlocal) presented above, is capable of predicting the nonlinear ring structure of the transverse beam profile and phase associated with the beam vortex states (non zero eigenvalue of the orbital angular momentum). Moreover, such approach seems to be suitable for plasma-based beam focusing (plasma lens) to nano-sized beam spots.


%

\end{document}